\documentclass{article}

\usepackage{amsmath}
\usepackage{amssymb}
\usepackage{mathtools}
\usepackage{amsthm}

\usepackage{macros}
\usepackage[parfill]{parskip}
\usepackage[]{natbib}
\usepackage{xr-hyper,refcount}
\usepackage{graphicx}
\usepackage[utf8]{inputenc} 
\usepackage{url}            
\usepackage{booktabs}       
\usepackage{amsfonts}       
\usepackage{nicefrac}       
\usepackage{microtype}      
\usepackage[parfill]{parskip}
\usepackage{algorithm,algorithmic}
\usepackage{amsmath,amsthm,bbm}
\usepackage{mathtools}
\usepackage{cases}
\usepackage{comment}
\usepackage{subcaption}
\usepackage{algorithm,algorithmic}
\usepackage{color}
\usepackage{appendix}
\usepackage{xspace}
\usepackage{enumitem}
\usepackage[english]{babel}
\usepackage{authblk}
\usepackage{mathtools}
\usepackage{enumitem}
\usepackage{booktabs}
\usepackage[]{xcolor}
\usepackage{verbatim}
\usepackage{fullpage}
\usepackage[hidelinks]{hyperref}
\definecolor{OliveGreen}{rgb}{0,0.6,0}

\usepackage[normalem]{ulem}

\DeclareMathOperator*{\argmax}{arg\,max}

\title{A Taxonomy of Human and ML Strengths in Decision-Making to Investigate Human-ML Complementarity}

\author{Charvi Rastogi\footnote{indicates equal contribution. Corresponding authors: CR (\href{mailto:crastogi@cs.cmu.edu}{\texttt{crastogi@cs.cmu.edu}}) and LL (\href{mailto:leqil@cs.cmu.edu}{\texttt{leqil@cs.cmu.edu}}).}\;\,}
\author{Liu Leqi$^{*}$}
\author{Kenneth Holstein}
\author{Hoda Heidari}

\affil{School of Computer Science, Carnegie Mellon University}

\date{}

\begin{document}
\maketitle

\begin{abstract}

Hybrid human-ML systems increasingly make consequential decisions in a wide range of domains. These systems are often introduced with the expectation that the combined human-ML system will achieve \textit{complementary performance}, that is, the combined decision-making system will be an improvement compared with either decision-making agent in isolation. However, empirical results have been mixed, and existing research rarely articulates the sources and mechanisms by which complementary performance is expected to arise. Our goal in this work is to provide conceptual tools to advance the way researchers reason and communicate about human-ML complementarity. Drawing upon prior literature in human psychology, machine learning, and human-computer interaction, we propose a taxonomy characterizing distinct ways in which human and ML-based decision-making can differ. In doing so, we conceptually map potential mechanisms by which combining human and ML decision-making may yield complementary performance, developing a language for the research community to reason about design of hybrid systems in any decision-making domain. To illustrate how our taxonomy can be used to investigate complementarity, we provide a mathematical aggregation framework to examine enabling conditions for complementarity. Through synthetic simulations, we demonstrate how this framework can be used to explore specific aspects of our taxonomy and shed light on the optimal mechanisms for combining human-ML judgments.

\end{abstract}

\section{Introduction}\label{sec:intro}

In recent years, we have witnessed a rapid growth in the deployment of machine learning (ML) models in decision-making systems across a wide range of domains, including healthcare~\citep{patel2019human, rajpurkar2020chexaid, tschandl2020human, bien2018deep}, credit lending~\citep{bussman2021explainable, KRUPPA20135125}, criminal justice~\citep{angwin2016machine, kleinberg2018human}, and employment~\citep{raghavan2020mitigating, hoffman2017discretion}. For example, in the criminal justice system, algorithmic recidivism risk scores inform pre-trial bail decisions for defendants~\citep{angwin2016machine}. In credit lending, lenders routinely use credit-scoring models to assess the risk of default by applicants~\citep{KRUPPA20135125}. The excitement around modern ML systems facilitating high-stakes decisions is fueled by the promise of these technologies to tap into large datasets, mine the relevant statistical patterns within them, and utilize those patterns to make more accurate predictions at a lower cost and without suffering from the same cognitive biases and limitations as human decision-makers. Growing evidence, however, suggests that ML models are vulnerable to various biases~\citep{angwin2016machine} and instability~\citep{finlayson2018adversarial}. Furthermore, they often produce harmful outcomes in practice, given that they lack humans strengths such as commonsense reasoning abilities, cognitive flexibility, and social and contextual knowledge~\citep{alkhatib2021live,holstein2021designing, lake2017building, miller2019explanation}. These observations have led to calls for both human and ML involvement in high-stakes decision-making systems---with the hope of combining and amplifying the respective strengths of human thinking and ML models through carefully designed \emph{hybrid} decision-making systems. Such systems are common in practice, including in the domains mentioned above. %
 
Researchers have proposed and tested various hybrid human-ML designs, ranging from human-in-the-loop~\citep{russakovsky2015best} to algorithm-in-the-loop~\citep{de2020case, saxena2020human, brown2019toward, green2019principles} arrangements. However, empirical findings regarding the success and effectiveness of these proposals are mixed~\citep{holstein2021designing,lai2021towards}. Simultaneously, a growing body of theoretical work has attempted to conceptualize and formalize these hybrid designs~\citep{gao2021human, Bordt2020-ac} and study optimal ways of aggregating human and ML judgments within them~\citep{madras2018predict, mozannar2020consistent, wilder2020learning, keswani2021towards, Raghu2019algorithmic, okati2021differentiable, donahue2022human, steyvers2022bayesian}.

Much prior work has studied settings where the ML model outperforms the human decision-maker. These studies are frequently focused on tasks where there are no reasons to expect upfront that the human and the ML model will have complementary strengths~\citep{bansal2021most,guerdan2023ground,holstein2021designing,LurieMulligan2020}. For example, some experimental studies employ untrained crowdworkers on tasks that require extensive domain expertise, without which there is no reason to expect that novices would have complementary strengths~\citep{fogliato2021impact,LurieMulligan2020, rastogi2020deciding}. Other experimental studies are designed in ways that artificially constrain human performance---for instance, by eliminating the possibility that humans and ML systems have access to complementary information~\citep{guerdan2023ground}. Meanwhile studies on human-ML decision-making in real-world settings such as healthcare~\citep{tschandl2020human, patel2019human} sometimes demonstrate better human-ML team performance than either agent alone. However, the \emph{reasons} for complementary team performance are often left unexplained, where we define human-ML \emph{complementarity} as the condition in which a combination of human and ML decision-making outperforms\footnote{Complementary performance may present along any performance metric, and does not necessarily refer to accuracy.} both human- and ML-based decision-making in isolation. 

\begin{figure*}
    \centering
    \includegraphics[width=\textwidth]{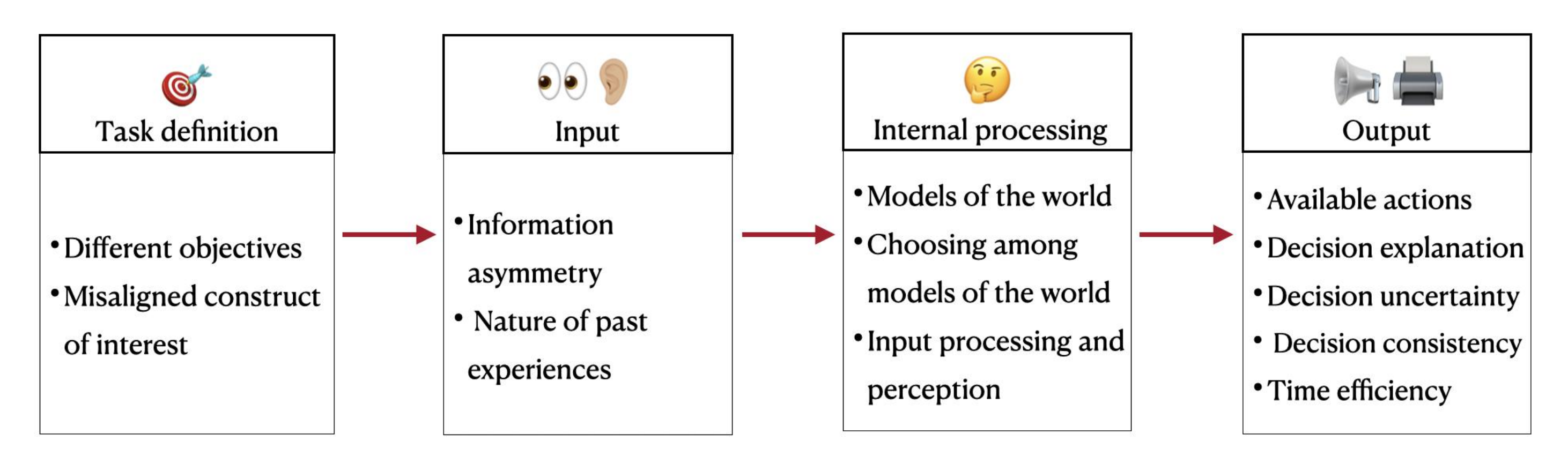}
    \caption{Proposed taxonomy of human and ML strengths \& weaknesses in decision-making divided into four parts of the decision-making process: Task definition, input, internal processing, and output.}
    \label{fig:taxonomy}
\end{figure*}

We argue, therefore, that there is a clear need to form a deeper, more fine-grained understanding of what types of human-ML systems exhibit complementarity in combined decision-making. To respond to this gap in the literature, we build a novel \emph{taxonomy} of relative strengths and weaknesses of humans and ML models in decision-making, presented in Figure~\ref{fig:taxonomy}. This taxonomy aims to provide a shared understanding of the causes and conditions of complementarity so that researchers and practitioners can design more effective hybrid systems and focus empirical evaluations on promising designs---by investigating and enumerating the distinguishing characteristics of human vs. ML decision-making upfront. Our taxonomy covers application domains wherein the decision at stake is solely based on \emph{predicting} some outcome of interest \citep{mitchell2018prediction}. Henceforth, we use the terms `prediction' and `decision' interchangeably. Some examples of predictive decisions are diagnosis of diabetic retinopathy~\citep{gulshan2016development}, predicting recidivism for pretrial decisions~\citep{dressel2018accuracy}, and consumer credit risk prediction~\citep{bussman2021explainable}.

To build our taxonomy of human-ML complementarity, we surveyed the literature on human behavior, cognitive and behavioral sciences, as well as psychology to understand the essential factors across which human and ML decision-making processes differ. Following traditions in cognitive science and computational social science~\citep{lake2017building, marr1977understanding}, we understand human and ML decision-making through a computational lens. Our taxonomy maps distinct ways in which human and ML decision-making can differ (Section~\ref{sec:taxonomy}). 

To illustrate how our taxonomy can be used to investigate when we can expect complementarity in a given setting and what modes of human-ML combination will help achieve it, we present a mathematical framework that captures each factor in the taxonomy. In particular, we formalize an optimization problem for convex combination of human and ML decisions. This problem setup establishes a pathway to help researchers explore which characteristics of humans and ML models have the potential to foster complementary performance. To categorize different types of complementarity, we propose quantitative measures of complementarity, designed to capture two salient modes of human-ML collaboration in the literature: routing (or deferral) and communication-based collaboration. To demonstrate the use of our taxonomy, the optimization problem setup, and the associated metrics of complementarity, we simulate optimal human-ML combinations under two distinct conditions: (1) human and ML models have access to different feature sets, (2) human and ML models have different objective functions. By comparing optimal aggregation strategies under these conditions, we gain critical insights regarding the contribution of each decision-making agent towards the optimal combined decision. This informs the effective design of human-ML partnerships under these settings for future research and practice. Taken together, this work highlights that combining human-ML judgments should leverage the unique strengths and weaknesses of each entity, as different sources of complementarity impact the extent and nature of performance improvement achievable through human-ML collaboration.

In summary, this paper contributes a unifying taxonomy and formalization for human-ML complementarity. Our taxonomy characterizes major differences between human and ML predictions, and our optimization-based framework formally characterizes optimal aggregation of human and machine decisions under various conditions and the type of complementarity that produces optimal decisions. With these contributions, we hope to provide a common language and an organizational structure to inform future research in this increasingly important space for human-ML combined decision-making.

\section{Methodology for Designing the Taxonomy}
\label{sec:background}

To investigate the potential for complementarity in human-ML combined decision-making, we need to understand the respective strengths and drawbacks of the human decision-maker and the ML model in the context of the application. For instance, it has been observed that while ML models draw inferences based on much larger bodies of data than humans could efficiently process~\citep{jarrahi2018future}, human decision-makers bring rich contextual knowledge and common sense reasoning capabilities~\citep{holstein2021designing, miller2019explanation, lake2017building} to the decision-making process, which ML models may be unable to replicate. Thus, we develop a taxonomy for human-ML decision-making that accounts for broad differences between human decision-makers and machine learning, encompassing applications with predictive decision-making.

  To inform this taxonomy, we draw from existing syntheses in human psychology, machine learning, AI, and human-computer interaction to understand distinguishing characteristics that have been observed between human decision-makers and ML in the context of predictive decision-making. In cognitive science,~\citet{lake2017building} review major gaps between human and ML capabilities by synthesizing existing scientific knowledge about human abilities %
that have thus far defied automation. In management science,~\citet{shrestha2019organizational} identify characteristics of human and ML decision-making along four key axes: decision space, process and outcome interpretability, speed, and replicability, and discuss their combination for organizational decision-making. In human-computer interaction,~\citet{holstein2020conceptual} conceptually map distinct ways in which humans and AI can augment each others' abilities in real-world teaching and learning contexts. More recently,~\citet{lai2021towards} surveyed empirical studies on human-AI decision-making to document trends in study design choices (e.g., decision tasks and evaluation metrics) and empirical findings. We draw upon this prior literature to summarize key differences in human and ML-based predictive decision-making across multiple domains, with an eye towards understanding opportunities to combine their strengths.

\textbf{Computational lens.} Our taxonomy takes a computational perspective towards analysing human and ML decision-making. As with any modeling approach or analytic lens, computational-level explanations are inherently reductive, yet are often useful in making sense of complex phenomena for this very reason. Computational-level explanations often provide an account of a \textit{task} an agent performs, the \textit{inputs} that the agent takes in, the ways in which the agent \textit{perceives and processes} these inputs, and the kinds of \textit{outputs} that the agent produces. Accordingly, our taxonomy is organized into four elements: (1) task definition, (2) input, (3) internal processing, and (4) output.

We now provide mathematical notation to clearly express the computational perspective of decision-making in our taxonomy. Formally, the agent's decision-making setting is specified by a feature space $\statespace$, an action space $\actionspace$, and the  space of observed outcomes, $\outcomespace$. At a high-level, the agent perceives an instance $\vecx \in \statespace$, chooses to take an action $\action \in \actionspace$ based on its relevant prior knowledge and experiences, and observes an outcome $\outcome\in\outcomespace$ as a result. To emphasize that the outcome $\outcome$ is influenced by $\vecx$ and $\action$, we slightly abuse the notation to denote the outcome of action $\action$ on instance $\vecx$ as $\outcome(\vecx, \action)$. We consider the agent's perception of an instance $\vecx$ to be denoted by $\featuremap(\vecx)$, where $\featuremap:\statespace \to \statespace$. Next, the agent's prior knowledge and relevant experiences are assumed to be encompassed in a set $\dataset$.
The goal of the decision-making agent is to choose a policy, $\policy:\statespace \to {\actionspace}$, 
in the space of feasible policies $\policyspace$,
such that $\policy$ leads to favorable overall outcome quality, measured by an evaluation function $\evalf$. Here $\evalf$ takes in a policy and outputs a real number. For instance, the expected outcome of a policy is a common choice for $\evalf$.
Finally, the agent chooses their optimal policy $\opolicy$ using their optimization process $\OPT$ for choosing among feasible policies that lead to favorable $\evalf$ values. Using the categorization and mathematical formalization above, and drawing upon relevant background literature as presented in this section, we now provide our taxonomy for relative human and ML strengths.

\section{A Taxonomy of Human and ML Strengths \& Weaknesses in Decision-Making}
\label{sec:taxonomy}
In this work, we consider two decision-making agents, the human and the ML model denoted respectively by $\h{}$ and $\m{}$. Building upon the notation in Section~\ref{sec:background}, we denote the feature space available to each agent by $\statespace_\h, \statespace_\m$ correspondingly, where $\statespace_\h, \statespace_\m \subseteq \statespace$. Similarly, for each variable introduced for our decision-making setting in the previous section, we consider a human version and a ML version, denoted by subscript $\h$ and $\m$ respectively. We now present our taxonomy, visually represented in Figure~\ref{fig:taxonomy}.

\subsection{Task Definition}
We now describe the distinguishing characteristics that have been observed in the definition of the decision-making task used by the human and the ML model. 
\begin{itemize}[leftmargin=*]

    \item \textbf{Objective.} 
   Most machine learning models aim to only optimize the expected performance, e.g., minimize the expected loss for supervised learning models and maximize the expected cumulative rewards for reinforcement learning models. 
   While recent research has explored ways to build models with respect to a more diverse set of objectives, including different risk measures \citep{liu2019human,khim2020uniform},
    fairness definitions~\citep{chouldechova2020snapshot} and interpretability notions~\citep{lipton2018mythos, miller2019explanation}, it is often difficult or impractical to encode all aspects of the objectives that a human decision-maker would aim to optimize~\citep{kleinberg2018human}.  Using our notation, this is expressed as $\evalf_\h \neq \evalf_\m$. For example, 
    when making a lending decision, in addition to considering various risk factors, bankers may also care about aspects such as maintaining their relationships with clients and specific lending practices in their organization~\citep{tronnberg2014lending}. 
    
    \item \textbf{Misaligned construct of interest.} ML models deployed in social contexts often involve theoretical constructs that are not directly observable in the data, such as socioeconomic status, teacher effectiveness, and risk of recidivism, which cannot be measured directly. Instead they are inferred indirectly via proxies: measurements of properties that are observed in the data available to a model. The process of defining proxy variables for a construct of interest necessarily involves making simplifying assumptions, and there is often a considerable conceptual distance between ML proxies and the ways human decision-makers think about the targeted construct~\citep{green2021algorithmic,guerdan2023ground,jacobs2021measurement,kawakami2022improving}. In other words, $\outcome_\h(\vecx, \action) \neq \outcome_\m(\vecx, \action)$. \citet{jacobs2021measurement} argue that several harms studied in the literature on fairness of socio-technical systems are direct results of the mismatch between the construct of interests and the inferred measurements. 
     For example, \citet{obermeyer2019dissecting} examined racial biases in an ML-based tool used in hospitals. They found that the use of an indirect proxy (healthcare costs incurred by a patient) to predict patients' need for healthcare contributed to worse healthcare provision decisions for black versus white patients. In this example, although the proxy used (the monetary cost of care) was conveniently captured in available data, it differs significantly from the way healthcare professionals conceptualize patients' actual need for care.
    
\end{itemize}

\subsection{Input} 

We now describe the distinguishing characteristics observed in the inputs used by humans and ML models.
\begin{itemize}[leftmargin=*]
    \item \textbf{Access to different information.} From the input perspective, in many settings such as healthcare, criminal justice, humans and machines have access to both shared and non-overlapping information: $\statespace_\h \neq \statespace_\m$. This is because real-world decision-making contexts often contain features of importance that cannot be codified for ML. For example, a doctor can see the physical presentation of a patient and understand their symptoms better, since this information is hard to codify and provide to the machine. Similarly, a judge learns about the predisposition of the defendant through interaction~\citep{kleinberg2018human}. This phenomena is also referred to as unobservables~\citep{holstein2023toward} and information asymmetry~\citep{hemmer2022effect} in the literature on human-ML complementarity.
    
\item \textbf{Nature of past experiences.}
The nature of embodied human experience over the course of a lifetime differs substantially from the training datasets used by modern ML systems: $\dataset_\h \neq \dataset_\m$. For example, ML models are often trained using a large number of prior instances of a specific decision-making task, but for each instance, the training data contains a fixed and limited set of information. This often does not reflect the richness of human experience. Humans make their decisions with reference to a lifetime of experiences across a range of domains, and it is difficult to explicitly specify the information they take into account.
By contrast, ML models may learn from training data that comprise narrow slices from a vast number of human decision-makers' decisions, whereas humans typically learn only from their own experiences or from a small handful of other decision-makers. 
\end{itemize}

\subsection{Internal Processing}
\label{internalproc}

We now describe the distinguishing characteristics 
of the internal processes used by humans and ML systems.

\begin{itemize}[leftmargin=*]
    \item 
    \textbf{Models of the world.}
    As is comprehensively overviewed  in~\citet{lake2017building}, humans rely upon rich mental models and ``theories'' that encode complex beliefs about causal mechanisms in the world, not just statistical relationships. This results in humans having a different set of models of the world than those embodied by ML models: $\policyspace_\h \neq \policyspace_\m$. For example, starting from an early age, humans develop sophisticated systems of beliefs about the physical and social worlds (intuitive physics and intuitive psychology), which strongly guide how they perceive and make decisions in the world. In contrast to modern ML systems, humans' mental models tend to be compositional and causal.
    In turn, these strong prior beliefs about the world can enable humans to learn rapidly in comparison to modern ML systems, and to make inferential leaps based on very limited data (e.g., one-shot and few-shot learning) \citep{gopnik2012reconstructing,lake2017building,tenenbaum2011grow}.
    On the other hand, the model class of the machine decision-maker has a more mathematically tractable form---whether it is a class of parametric or non-parametric models~\citep{friedman2017elements}. Although when designing these models such as neural networks, 
    researchers commonly encode domain knowledge through the data and the model architecture, most machine learning models still suffer from distribution shift~\citep{quinonero2009dataset} and lack of interpretability~\citep{gilpin2018explaining}, and require large sample sizes.

    \item \textbf{Input processing and perception.} 
 The ways decision-makers perceive inputs is informed by their models of the world \citep{gentner2014mental,holstein2020conceptual}. 
 Following research in human cognition and ML, we highlight three sources of variation in input perception: (1) differences in mental/computational capacity, (2) differences in human versus machine biases, and (3) tendencies towards causal versus statistical perception. Here the first implies $\featuremap_\h \neq \featuremap_\m$ and the remaining two indicate $\policy_\h \neq \policy_\m$. For instance, compared with ML systems, humans demonstrate less capacity to perceive small differences in numerical values~\citep{Amitay2013HumanDM, FINDLING2021computation}. Furthermore, both humans and ML systems can bring in both adaptive and maladaptive biases, based on their experiences and models of the world, which in turn shape the ways they process and perceive new situations~\citep{Fitzgerald2017ImplicitBI, wistrich2017implicit,kleinberg2018human,gentner2014mental}. However, in some cases humans and ML systems may have complementary biases, opening room for each to help mitigate or compensate for the other's limitations~\citep{holstein2020conceptual,tan2018investigating}. Finally, research on human cognition demonstrates that humans are predisposed to perceiving causal connections in the world, and drawing causal inferences based on their observations and interactions in the world~\citep{gopnik2012reconstructing,lake2017building}. While these abilities can sometimes be understood by analogy to the kinds of statistical learning that most modern ML systems are based upon~\citep{tenenbaum2011grow}, other aspects of human causal cognition appear to be fundamentally different in nature~\cite{lake2017building}. As with bias, these abilities can be a double-edged sword. In some scenarios, human causal perception may lead to faulty inferences based on limited data. By contrast, ML systems will sometimes have an advantage in drawing more reliable inferences based on statistical patterns in large datasets. In other settings, human causal perception can help to overcome limitations of ML systems. For example, in many instances, human decision-makers have been observed to be better than ML systems at adapting to out-of-distribution instances, through the identification and selection of causal features for decision-making~\citep{lake2017building}.

    \item \textbf{Choosing among models of the world.}
    Given the task definition, models of the world, and data, ML models differ from humans in searching for the model that optimizes their objective: $\OPT_\h \neq \OPT_\m$. Modern ML models (e.g., neural networks) are commonly learned using first-order methods and may require a huge amount of computational resource due to the size of the models~\citep{bottou2010large}. 
    On the other hand, humans may employ heuristics that can be executed in a relatively short amount of time~\citep{simon1979rational}. These simple strategies may have advantages over more complex models when the inherent uncertainty in the task is high. 
    For a more comprehensive review on when and how such heuristics may be more preferable, we refer readers to~\citet{kozyreva2021interpretation}. 
\end{itemize}

\subsection{Output} We now describe the distinguishing characteristics of the outputs generated by humans and ML systems. 

        \begin{itemize}[leftmargin=*]

        \item \textbf{Available actions.} In real-world deployment settings, the set of possible decisions or actions available to ML models versus humans can be different: $\actionspace_\h \neq \actionspace_\m$. For example, in the context of K-12 education, ML-based tutoring software may be able to provide just-in-time hints to students, to help struggling students them with math content. Meanwhile, although a human teacher working alongside this software in the classroom has limited time to spend with each student, they can take a wider range of actions to support students, such as providing emotional support or helping students with prerequisite content that lies outside of the software's instructional repertoire~\citep{holstein2020conceptual}. Similarly, in the context of ML-assisted child maltreatment screening, a model may only be able to recommend that a case be  investigated or not investigated, based on the information that is currently available. By contrast,~\citet{kawakami2022improving} report that human call screeners may take actions to gather additional information as needed, e.g. by making phone calls to other stakeholders relevant to a case.
        
        \item  \textbf{Explaining the decision.} 
        Humans and ML have differing abilities in communicating the reasoning behind their decisions. 
        There has been extensive research in explainability (XAI) and interpretability for ML~\citep{adadi2018peeking}. Research in cognitive and social psychology observes that humans are generally better than ML algorithms at generating coherent explanations that are meaningful to other humans. Furthermore,~\citet{miller2019explanation} argues that XAI research should move away from imprecise, subjective notions of ``good'' explanations and instead focus on reasons and thought processes that people apply for explanation selection. They find that human explanations are contrastive, selected in a biased manner, and most importantly they are social and contextual.  On the other hand, humans’ explanations may not have a correspondence to their actual underlying decision processes~\citep{nisbett1977telling}, whereas with ML models we can always trace the precise computational steps that led to the output prediction~\citep{hu2019optimal}.
        \item \textbf{Uncertainty communication.} With increasing research in uncertainty quantification for machine learning, new methods have been devised for calibrating a ML model's uncertainty in its prediction~\citep{abdar2021review}. Moreover, methods have been developed to decompose the model uncertainty into aleatoric uncertainty and epistemic uncertainty~\citep{Hullermeier2021aleatoric}, where aleatoric uncertainty signifies the inherent randomness in an application domain and cannot be reduced, and epistemic uncertainty, also known as systematic uncertainty, signifies the uncertainty due to lack of information or knowledge, and can be reduced. However, these uncertainty quantification methods may not necessarily be well-calibrated~\citep{abdar2021review}, and are an active research direction. Meanwhile, human decision-makers also find it difficult to calibrate their uncertainty or their confidence in their decisions~\citep{brenner2005modeling}, and tend to output discrete decisions instead of uncertainty scores. Moreover, different people have different scales for uncertainty calibration~\citep{zhang2012ubiquitous}. 
    \item \textbf{Output consistency.} We define a given decision-maker to have a consistent output when they always produce the same output for the same input. Therefore, we consider the inconsistency in decisions that are based on factors independent of the input, we call them extraneous factors. Some examples of extraneous factors are the time of the day, the weather, etc. Research in human behavior and psychology has shown that human judgments show inconsistency~\citep{kahneman2016noise}. More specifically, there is a positive likelihood of change in outcome by a given human decision-maker given the exact same problem description at two different instances. Within-person inconsistency in human judgments has been observed across many domains, including medicine~\citep{koran1975reliability, Kirwan1983ClinicalJI}, clinical psychology~\citep{little1961confidence}, finance and management~\citep{kahneman2016noise}. This form of inconsistency is not exhibited by standard ML algorithms.\footnote{There exists the special case of randomized models, we consider these outside the scope of our work and, further note that these models can be directly mapped to deterministic models with decision-based thresholds.} %

   \item \textbf{Time efficiency.} In many settings, ML models can generate larger volumes of decisions in less time than human decision-makers.
   In addition to potentially taking more time per decision, humans often have comparatively scarce time for decision-making overall.
\end{itemize}

\section{Investigating the Potential for Human-ML Complementarity} 
\label{sec:optimization}

To understand how the differences in human and machine decision-making result in complementary performance, we formulate an optimization problem to aggregate the human and the ML model outcomes. The key motivation here is to use information available about human and ML decision-making (in the form of historical data or decision-making models) to understand the potential for complementarity in human-ML joint performance. Specifically, this optimization problem outputs the optimal convex combination of the two decision-makers' outputs wherein the aggregation mechanism represents the best that the human-ML joint decision-making can achieve in our setting.

In our decision-making setting, as mentioned in Section~\ref{sec:background}, we consider a feature space $\statespace$, an action space $\actionspace$ and an outcome space $\outcomespace$. Given a problem domain, {the} goal is to combine the two decision-makers policies to find a joint policy denoted by $\opolicy:\statespace \to {\actionspace}$  that  maximizes the overall quality of the decisions based on evaluation function, $\evalf{}$, 
\begin{align}\label{eq:general-obj}
    \opolicy \in 
    \argmax_{\policy \in \Pi} \; \evalf(\policy).
\end{align} 
We note that the overall evaluation function $\evalf$ for the joint policy $\policy$ may be different from that used by the human $\evalf_\h$ or the ML model $\evalf_\m$. We assume the joint policy is obtained by combing human and machine policies $\policy_\h$ and $\policy_\m$ over $\totalnum{}$ number of instances
through an aggregation function. We consider the outcome space to be scalar $\outcomespace \subseteq \Real$. 
Given $\policy_\h \in \policyspace_\h$, 
$\policy_\m \in \policyspace_\m$,  for an instance $\vecx_i$ where $i \in [ \totalnum{}]$, 
the joint policy $\pi \in \policyspace$ is given by 
\begin{align}\label{eq:aggregation-weighting}
   \policy(\vecx_i) = 
    \whi \policy_\h(\vecx_i) + \wmi \policy_\m(\vecx_i),
\end{align}
for some weights $\whi, \wmi \in [0,1]$ and $\whi + \wmi = 1$ for all $i \in  [\totalnum{}]$.
Here note that we assume that the joint decision $\policy(\vecx_i)$
is a convex combination of the individual decisions 
$\policy_\h(\vecx_i)$ and $\policy_\m(\vecx_i)$.
This {assumption arises naturally to }ensure that the joint decision lies between the human's and machine's decision. For a decision-maker (say human), the weight assigned for instance $i$, $\whi$ indicates the amount of contribution from them towards the final decision:
when $\whi=0$, 
the joint decision does not follow human's decision at all on instance $\vecx_i$, while $\whi=1$ indicates that their decision is followed entirely. {For the optimal policy $\opolicy$ defined in \eqref{eq:general-obj}, its corresponding optimal weights are denoted by $\owhi$ and $\owmi$.}

Several existing works on human-ML combination for decision-making, such as~\citet{donahue2022human, Raghu2019algorithmic, mozannar2020consistent, gao2021human} are subsumed by our convex combination optimization setup. Particularly, our aggregation mechanism captures two {salient} modes:
(1) The mode where an instance is routed to either the human or the ML decision maker, also known as deferral. This is represented by $\whi,\wmi \in \{0,1\}$ for all $i \in [\totalnum]$.
(2) The mode where a joint decision {lying between the human and the ML decision} is applied to each instance. This is represented by $\whi, \wmi \in (0,1)$ for all $i \in [\totalnum]$.

\subsection{Metrics for Complementarity}
\label{sec:metrics}

The proposed aggregation framework is a way to inspect the extent of complementarity in human-ML joint decision-making. {Recall that, based on our definition,}
The joint policy $\policy$ defined in \eqref{eq:aggregation-weighting} exhibits complementarity  if and only if  $$\evalf(\policy) > \max\{\evalf(\policy_\h), \evalf(\policy_\m)\}.$$
{Although this criterion provides a binary judgment on whether complementarity exists in a particular joint decision-making setting, 
it cannot be used to compare the amount of potential for complementarity in different settings. For instance, between two settings where machine can improve the performance of the human decision-maker on one instance versus on all instances, one may say that there is more complementarity exhibited in the second setting.} Further, it does not distinguish between the two salient modes of combination defined above, where the second mode may require more interaction between the human and the machine decision-maker. So, to investigate the potential for complementarity in different settings more thoroughly, we {introduce} metrics for quantifying the complementarity between the human and ML
decision-maker. 

Specifically, we introduce the notion of within- and across-instance complementarity to represent the two modes of combination where for an instance $\vecx_i$, we either have only one of human or ML contributing to the final decision ($\wmi = 1 \text{ or } \whi = 1$), or both decision-makers contributing to the final decision partially ($\wmi > 0 \text{ and } \whi > 0$). These two types of combinations represent two {ways of achieving} complementarity. 
{In the first one}, there is no complementarity within a single {task} instance, since only the human or the ML model decision gets used. In this scenario, if the human and ML model provide the final decision for different instances of the task, we call this \emph{across-instance complementarity}. {In the second one}, {if} both human and ML model contribute to the same instance $\vecx_i$, we call this \emph{within-instance complementarity}. These two metrics help distinguish between different instance allocation strategies in human-ML teams described in~\cite{roth2019function}. 
Formally, given the weights assigned to the two agents in the final decision, we define the two metrics as follows:
\begin{itemize}[leftmargin=*]
    \item  \textbf{Across-instance complementarity} quantifies the variability of the human (or the machine) decision-maker's contribution to the final decision across all task instances. Therefore, we define it as the variance of the weights assigned, written as
    \begin{align}
    \begin{split}
                \across{}(w_\m, w_\h) &\coloneqq \frac{1}{\totalnum} \sum_{i=1}^n \left(\wmi- \frac{1}{\totalnum} \sum_{i=1}^n \wmi \right)^2  \\
                &= \frac{1}{\totalnum} \sum_{i=1}^n \left(\whi- \frac{1}{\totalnum} \sum_{i=1}^n \whi \right)^2.
        \label{eq:across}
        \end{split}
    \end{align}
    The equality follows directly using the constraint $\wmi + \whi = 1$. In case of no variability across instances, that is if for both decision-makers, we have $\wmi$ (or $\whi$) to be a constant for all $i \in [\totalnum]$, then $\across{}(w_\m, w_\h) = 0$. The notion of across-instance complementarity is shown by works on decision deferral or routing including~\citet{mozannar2020consistent, madras2018predict}.
   \item  \textbf{Within-instance complementarity} quantifies the extent of collaboration between the two decision-makers on each individual task instance.
    Formally, we define 
    \begin{align}
    \within{}(w_\m, w_\h) \coloneqq 1 - \frac{1}{\totalnum} \sum_{i=1}^n \left(\whi  - \wmi \right)^2.
    \label{eq:within}
\end{align}
  Importantly, the definition of within-instance complementarity satisfies some key properties: $\within{}(w_\h, w_\m) $  is maximized at $\whi = \wmi = 0.5$ and minimized at $\whi \in \{0,1\}$ for all $i \in [\totalnum]$. Thus, it is maximized when each decision-maker contributes equally and maximally to a problem instance and minimized when there is no contribution from one of the decision-makers. Further, it increases monotonically as $\whi$ and $\wmi$ get closer to each other in value, that is the two decision-makers' contributions to the final decision get closer to half. This notion of complementarity is demonstrated in several works including~\citet{patel2019human,tschandl2020human}. 
\end{itemize}

\noindent To have a better grasp on the above two metrics, and to understand the importance of each metric in measuring complementarity, we provide some demonstrative examples. Consider a simple setting with $\statespace = \{\vecxinstantiation_1, \vecxinstantiation_2, \vecxinstantiation_3, \vecxinstantiation_4\}$ where each instance is equally likely, that is, $\mathbb{P}(\vecx = \vecxinstantiation_i) = 1/4$ for all $i \in [4]$. The values of the two metrics under different aggregation weights are given below: 

\begin{enumerate} 
    \item If $\weight_\h^{(1)} = \weight_\h^{(2)} = \weight_\h^{(3)} = \weight_\h^{(4)}=0$, then $\within{} = 0$, and $\across{} =0$.%
    \item If $ \weight_\h^{(1)} = \weight_\h^{(2)} = 0, \weight_\h^{(3)} = \weight_\h^{(4)} = 1$, then $\within{}= 0$, and  $\across{} = 0.25$.%
    \item If $\weight_\h^{(1)} = \weight_\h^{(2)} = \weight_\h^{(3)} = \weight_\h^{(4)} = 0.3$, then $\within{} = 0.84$, and $\across{} = 0$.%
\end{enumerate}

\noindent We note that although the second example has $\within{}= 0$ and $\across{} >0$, which is the opposite of the third example, both the  examples demonstrate complementarity. This shows that each metric introduced captures aspects of human-ML complementarity that is not captured by the other metric.

\section{Synthetic Experiments to Illustrate Complementarity}
\label{sec:simulations}

In this section, we illustrate how our proposed framework can be used {to investigate the extent and nature of complementarity} via simulations. These simulations utilize human and ML models learned from data, where the two decision-makers have different access of information or {they pursue} different objectives. By quantifying the extent of different types of complementarity {(i.e., within-instance and across-instance)}, we show how the proposed taxonomy and complementarity metrics can guide the research and practice of {hypothesizing about and testing for} complementarity with different types of human and ML decision-makers. To conduct these simulations, we choose specific aspects from our taxonomy in Section~\ref{sec:taxonomy} and measure complementarity in the presence of corresponding differences between the human and the ML model. We note that these simulations are meant to be an illustrative and not exhaustive exploration of human-ML complementarity {conditions} that can be explored using the taxonomy.

\begin{figure*}[t]
    \centering
    \includegraphics[width=\textwidth]{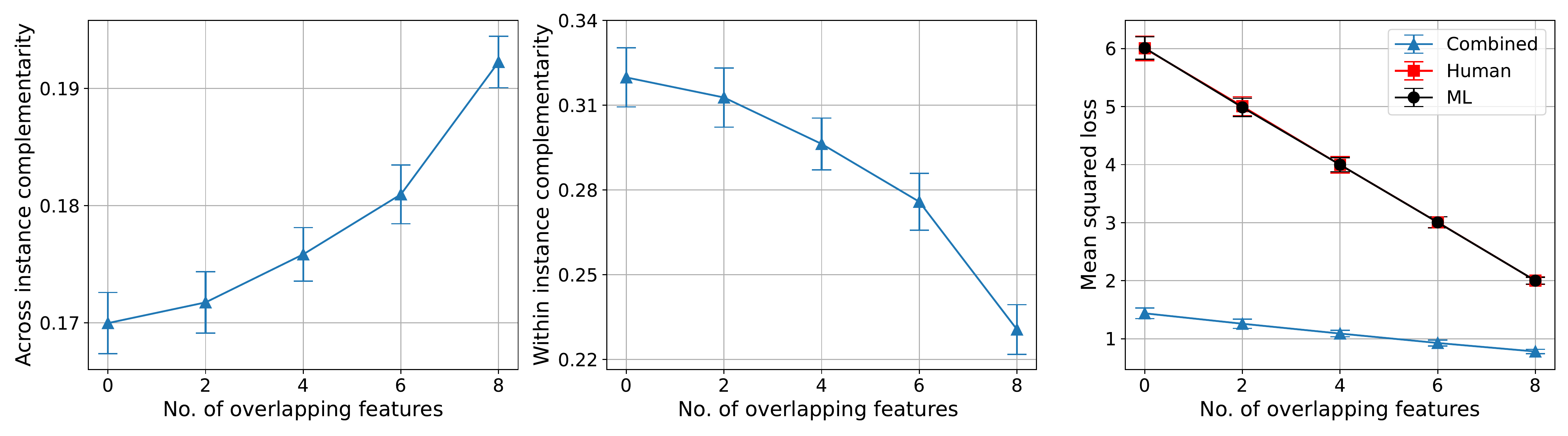}
    \caption{We plot the outcomes of Experiment I described in Section~\ref{sec:simulation-information}. The x-axis indicates the number of features that both the human and the ML model have access to. In each of the three figures, we plot an outcome metric for the optimal joint policy, namely across-instance complementarity~\eqref{eq:across}, within-instance complementarity~\eqref{eq:within} and mean squared loss of the policy compared to the target outcome $\target$. The markers show the mean value and the error bars indicate the standard deviation, based on 200 iterations. On the $x$-axis, we skip $x = 10$, as it is a straightforward setting where both the agents have access to all the features, so there is no complementarity, $\within{} = \across{} = 0$. Note that all three plots have different ranges on the $y$-axis, with $\across{}\in[0,0.25]$, $\within{}\in[0,1]$. To read these plots, we focus on relative values within plots, and not on absolute values across plots. We observe that  $\across{}$ increases while $\within{}$ decreases as the number of overlapping features increases. When the agents have no overlapping features $(x=0)$ the two agents have more likely to be equally benefitial for each decision leading to a higher within-instance complementarity. Meanwhile, when both have largely overlapping information $(x=8)$, the combination is more likely to show across-instance complementarity, the gains of going with the better decision-maker outweighing the possible gains from combination on each instance. 
   }
    \label{fig:overlap}
\end{figure*}

{\xhdr{Synthetic simulation setup}} We consider a linear model for the data generating process: 
the features $\vecx\in \mathbb{R}^\dimn$ are distributed as $\vecx \sim \mathcal{N}(0, \mathbf{I}_{\dimn \times \dimn})$; the target is given by $\target  = \vecx{}^\top \beta + \epsilon$ where $\beta = (1 \cdots 1) \in \mathbb{R}^\dimn$ and $\epsilon \sim \mathcal{N}(0, 1)$.
For any given instance $\vecx \in \mathbb{R}^\dimn$, 
both the ML  and human decision-maker make a prediction using their respective linear model, which serves as a decision.  We assume that the outcome for a given instance is determined by the squared loss incurred by the decision. For example, for the machine, given the true target $\target$ and the prediction $\policy_\m(\vecx)$, the outcome  is given by $\outcome = (\policy_\m(\vecx) - \target)^2$. 
The dimension of the features is chosen to be $\dimn=10$ for all simulations. 

In Section~\ref{sec:simulation-information}, we study how human-ML complementarity varies when the human and the ML model have different feature information available to them; and in Section~\ref{sec:simulation-objective}, the difference between the human and the machine arises via difference in objective functions for learning their respective policies. In the following simulations, we first use a training set of sample size $8,000$ to learn the respective {optimal} linear model for the human and the ML policy. Once the decision-makers' policies are learned, a separate testing set of size $2,000$ is used to compute and analyse the optimal aggregation weights. On this set, we measure and report the metrics of complementarity defined in Section~\ref{sec:metrics}. 

\subsection{Access to Different Feature Sets}\label{sec:simulation-information}

First, we consider the setting where the human and the machine decision-maker have different information available to them. This is a potential source of complementarity in human-ML joint decision-making as mentioned in our taxonomy in Section~\ref{sec:taxonomy} based on the input. To analyze the impact of information asymmetry on human-ML complementarity, we conduct synthetic experiments based on the general setup described at the beginning of Section~\ref{sec:simulations}. Additionally, we assume that the features available to the human and the ML model are denoted by $\vecx_\h \in \mathbb{R}^{\dimn_\h} $ and $\vecx_\m \in \mathbb{R}^{\dimn_\m}$ respectively, where $\dimn_\h$ and $\dimn_\m$ indicate the number of features available to the human and the machine respectively, with $\dimn_\h,\dimn_\m \le \dimn = 10$. Given the input information available to them, the human and the machine learn a policy using linear regression on the training data, given by $\policy_\h: \mathbb{R}^{\dimn_\h}\to \mathbb{R}$ and $\policy_\m: \mathbb{R}^{\dimn_\m}\to \mathbb{R}$ respectively. 
Using the optimization problem setup in \eqref{eq:general-obj} and \eqref{eq:aggregation-weighting}, we conduct simulations to analyse the amount and type of complementarity achieved by the combination of human and ML agents with different information. Consequently, we conduct two sets of experiments. 

\begin{figure*}[t]
    \centering
    \includegraphics[width=\textwidth]{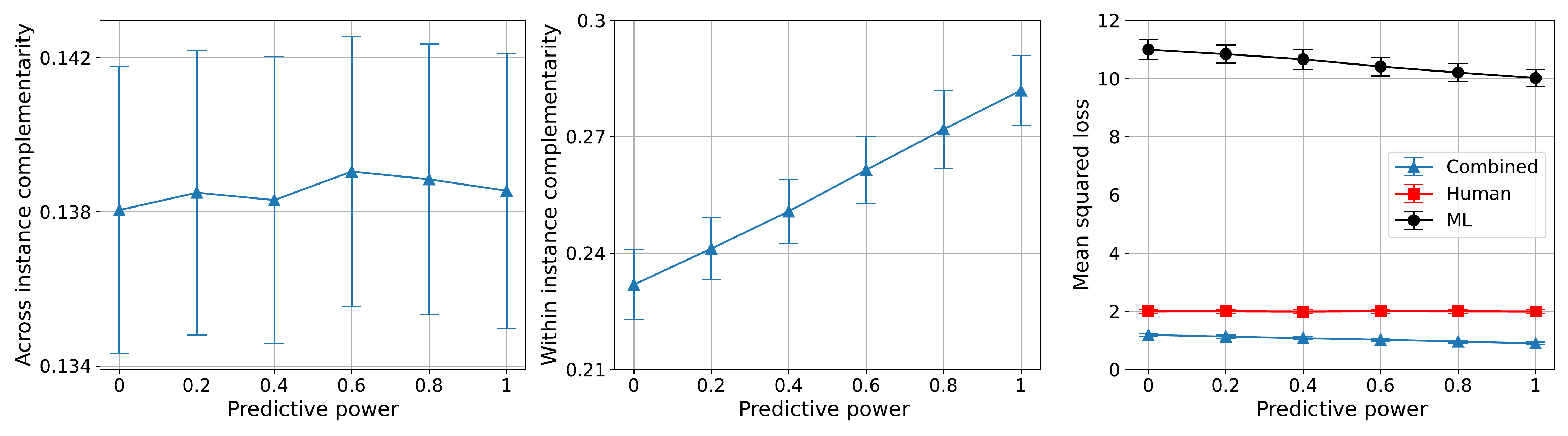}
    \caption{We plot the outcomes of Experiment II where the ML model has access to one feature and the human has access to the other nine features, described in Section~\ref{sec:simulation-information}. The $x$-axis indicates the predictive power of the feature $\vecx_\m$ that the machine has. In each of the three figures, we plot an outcome metric for the optimal joint policy, namely across-instance complementarity~\eqref{eq:across}, within-instance complementarity~\eqref{eq:within} and mean squared loss of the policy compared to the target outcome $\target$. The markers show the mean value and the error bars indicate the standard deviation, based on 200 iterations. Note that all three plots have different overall ranges on the $y$-axis, with $\across{}\in[0,0.25]$, $\within{}\in[0,1]$. To read these plots, we focus on relative values within plots. }
    \label{fig:alpha}
\end{figure*}

\paragraph{Experiment I.} We consider the setting where the human and ML have access to some common features and some non-common features as is typical of many real-world settings, as described in Section~\ref{sec:taxonomy}. Specifically, out of $\dimn = 10$ features in our setting, the human and the ML both have access to $z$ common features, and each has access to an additional $\frac{10-z}{2}$ features that only they can observe, where $z\in [\dimn]$. We plot the outcomes of this experiment in Figure~\ref{fig:overlap}, where the $x$-axis of each plot indicates $z$ (the degree of overlap between human and ML feature sets). Interestingly, we observe that while across-instance complementarity increases non-linearly with the number of overlapping features, within-instance complementarity decreases non-linearly. This suggests that when the two agents have access to many non-overlapping features, it would be important to use both the agents' decisions to come to a final decision on a given instance. On the other hand, in a setting with few overlapping features, the importance of collaboration on each instance reduces and it may be prudent to consider routing tasks to either the human or the machine for making the final decision. Furthermore, in the third plot, we observe that the combined decision has a strictly lower loss than either the human or the ML in isolation. Importantly, the gains achieved by the combined decision indicated by difference between the loss achieved by the individual agents and that by the combination is reducing as the number of overlapping features decreases. This suggests that depending upon the number of overlapping features and the resulting gain in accuracy, one may decide to forego joint human-ML decisions. We discuss this in more detail in Section~\ref{sec:discussion}.

\paragraph{Experiment II.} Next, we consider a setting where the human has access to nine of the features $\vecx_\h \in \mathbb{R}^9$ and the machine has access to the remaining tenth feature $\vecx_\m \in \mathbb{R}$. Within this setting, we simulate the types of information asymmetry identified in~\citet{holstein2023toward}. In this work on human-ML complementarity, the authors distinguish between non-overlapping features based on their ``predictive power'' which they define for any feature as the increase in training accuracy of a model as a result of including the feature. To simulate this, we vary the predictive power of the feature available to the ML model by introducing multiplicative random noise. Recall that $\target  = \vecx{}^\top \beta + \epsilon$ where $\beta = (1 \cdots 1) \in \mathbb{R}^\dimn$. Now, we define a variable $\alpha$ and let the data available to the ML model $\vecx_\m \in \mathbb{R}$ be based on $\alpha$ as: 
\begin{align}
  \vecx_\m &=  \begin{cases}
        \vecx_{10} \quad \text{if} \;\; \text{Binomial}(\alpha) = 1,\\
        0 \quad \;\;\;\;  \text{otherwise.}
    \end{cases}
\end{align}
In this manner, by varying $\alpha$ over the range $[0,1]$, we vary the predictive power of $\vecx_\m$. For $\alpha=0$ we have $\vecx_\m=0$ constantly, implying zero predictive power, and for $\alpha=1$ we have $\vecx_\m = \vecx_{10}$, implying the highest predictive power under the setting assumed. We show the outcomes of different complementarity measures under this setting in Figure~\ref{fig:alpha}. Observe that in the first plot, the across-instance complementarity does not change significantly with change in $\alpha$. The reasoning behind this is the human has a large majority of the features, thus having a high contribution in the final decision for all settings of $\alpha$.
On the other hand, within-instance complementarity increases linearly with $\alpha$, as increase in $\alpha$ implies that collaborating with the ML model on each instance will increase the predictive power of the overall policy. We also see that, as expected, the loss of the joint decision-maker improves as the predictive power increases.

\subsection{Different Objective Functions}\label{sec:simulation-objective}

\begin{figure}[t]
    \centering
    \includegraphics[width=.4\textwidth]{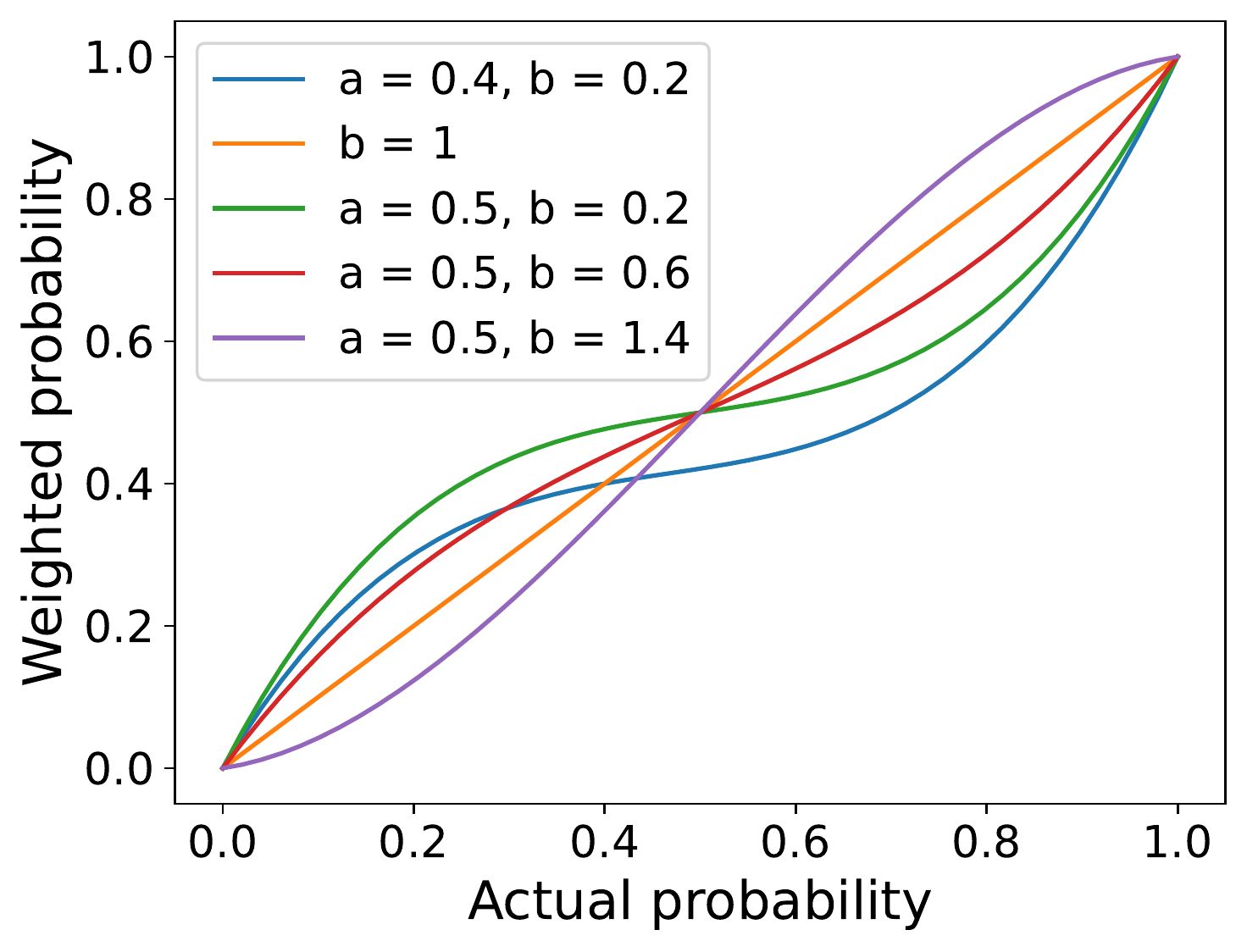}
    \caption{Examples of the probability weighting function used as the human's objective {based on CPT}. The $x$-axis specifies the actual probability and the $y$-axis indicates the perceived probability.  
    Parameter $a$ controls the fixed point and parameter $b$ controls the curvature of the function. 
    When $b < 1$, the probability weighting function has an inverted S-shape; when $b >1$, the function has an S-shape.
    }
    \label{fig:probability_weighting}
\end{figure}

In this setting, the human and ML decision-makers have different objectives,
which is a common source of {complementarity} in human and ML decision-making 
as noted in our taxonomy (Section~\ref{sec:taxonomy}).  
This may arise from the fact that ML models evaluate risks differently from humans. 
How agents evaluate the risks of an uncertain event is closely connected to how they perceive probabilities associated with this event. 
While ML models treat all probabilities according to their measured value, captured in their objective function as expected risk, humans tend to overweight small probabilities and underweight high ones, as suggested in Cumulative Prospect Theory (CPT)~\citep{tversky1992advances}. To capture this in our simulation, we model the human's objective function incorporating CPT as described in~\citet{leqi2019human}.

\begin{figure*}
    \centering
    \includegraphics[width=\textwidth]{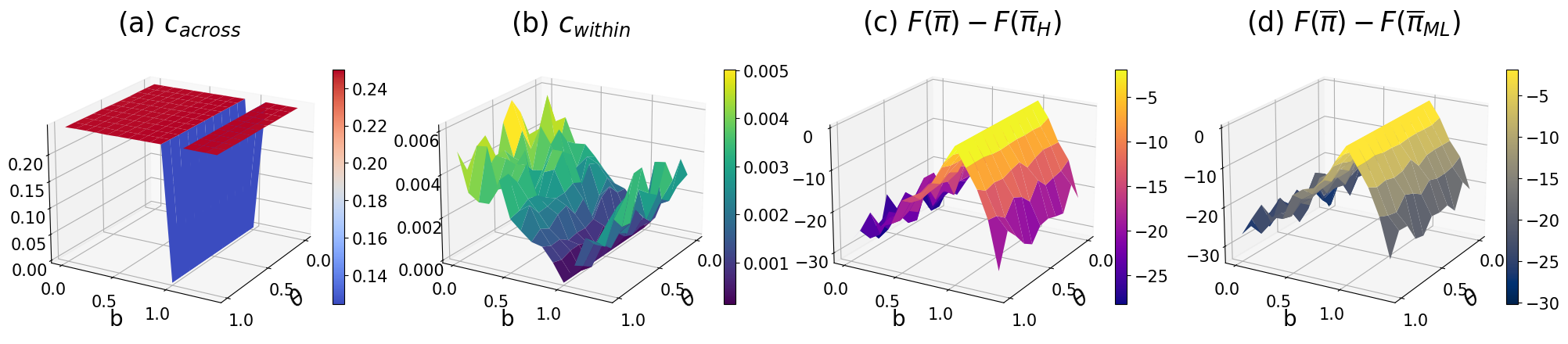}
    \caption{
    We plot the outcomes of the experiment where the ML and human have different objectives.
    For all plots, we set the probability weighting function parameter $a = 0.5$. 
    The $x$-axis gives the $b$ values, which specify the curvature of the probability weighting function; the $y$-axis gives the $\theta$ value, which specifies the overall objective function. 
    In the first two plots, the $z$-axis shows the across-instance complementarity $\across$ and within-instance complementarity $\within$, respectively. When $b=1$ (i.e., $\evalf_\h = \evalf_\m$), both $\across$ and $\within$ reach their lowest values. 
    We observe that $\across$ is high while $\within$ is low, 
    indicating that the final decision of each task instance is more likely to rely on a single agent. 
    In the last two plots, the $z$-axis shows $\evalf(\opolicy) - \evalf(\opolicy_\h)$ and $\evalf(\opolicy) - \evalf(\opolicy_\m)$, respectively. In both plots, the differences are below $0$, suggesting that the joint policy performs better compared to $\policy_\h$ and $\policy_\m$ under the overall objective function $\evalf$. All values are averaged across $5$ seeds.}
    \label{fig:obj}
\end{figure*}

More specifically, while the ML model's objective is to minimize the expected value of the squared error, $\evalf_\m(\policy_\m) = \frac{1}{n} \sum_{i=1}^n (\policy_\m(\mathbf{X}_i) - Y_i)^2$, 
the human's objective is to minimize $\evalf_\h(\policy_\h) =  \sum_{i=1}^n \frac{v_i}{n} (\policy_\h(\mathbf{X}_i) - Y_i)^2$ where $v_i$ reflects how humans overweigh and downweigh certain probabilities. As illustrated in Figure~\ref{fig:probability_weighting},
$v_i$ is parameterized by two parameters $a \in [0,1]$ and $b \in \mathbb{R}_+$ for specifying the fixed point and curvature of human's probability weighting function.\footnote{The exact form of $v_i$ is defined using the derivative of the probability weighting function shown in Figure~\ref{fig:probability_weighting}. More specifically, $v_i = \frac{3 - 3b}{a^2 - a + 1}  ( \frac{3 i^2}{n^2} - \frac{2(a+1)i}{n} + a) + 1$.} 
Notably, when $b=1$, the probability weighting function becomes the identity function and $v_i$ becomes $1$ for all $i \in [n]$, suggesting that $\evalf_\m = \evalf_\h$. 
For a more detailed explanation on the relation among the parameters $a, b$, the probability weighting function, and the factor $v_i$ in the objective function $\evalf_\h$, we refer the readers to~\citet{leqi2019human}[Section 3]. 
Lastly, we consider that the objective for the final decision balances between the human and the ML objective, defined as $\evalf(\policy) = \theta \evalf_\m(\policy) + (1-\theta) \evalf_\h(\policy)$ where $\theta \in [0,1]$ is a parameter controlling the overall objective function.
By varying parameters $\theta$, $a$ and $b$, we inspect how the difference in objective functions of the two agents and the joint decision affects the amount and type of complementarity that can be achieved in this setting.

As observed in Figure \ref{fig:obj} (c) and (d), 
the objective function differences $\evalf(\opolicy) - \evalf(\opolicy_\h)$ and $\evalf(\opolicy) - \evalf(\policy_\m)$ remain below $0$, suggesting that the learned joint policy outperforms both $\policy_\h$ and $\policy_\m$ under the overall objective function $\evalf$. 
For both across-instance complementarity $\across$ and within-instance complementarity $\within$, we find that when $b=1$, i.e., when the human and machine objectives are the same, their values are the lowest and are around $0$ (Figure \ref{fig:obj} (a) and (b)). 
{This is to be expected because} when the overall objective is the same as that of the human and the machine, there is no complementarity. 
When $b \neq 1$, $\across$ is relatively high while $\within$ is rather low,  suggesting that the optimal joint decision-maker does not need to rely on both agents for {making a decision on most instances}. 
Instead, a better form of collaboration between the human and the ML model is to defer each instance to one of the decision-makers. 
This is a {somewhat} unintuitive result 
since the overall objective function is a convex combination of the human's and the machine's, yet the final optimal decision is not. 
Importantly, this analysis shows evidence that we need to understand the mechanism of human-ML complementarity to inform how to {design the best aggregation mechanism}.
\section{Discussion}
\label{sec:discussion}
Our work contributes a deeper understanding of possible mechanisms for complementary performance in human-ML decision-making. Synthesizing insights across multiple research areas, we present a taxonomy characterizing potential complementary strengths of human and ML-based decision-making. Our taxonomy provides a pathway for reflection among researchers and practitioners working on human-ML collaboration to understand the potential reasons for expecting complementary team performance in their corresponding application domains. Our hope is that the research community will use this taxonomy to clearly communicate their hypotheses about the settings where they expect human-ML complementarity in decision-making.

Drawing upon our taxonomy, we propose a problem setup for optimal convex combination of the human and ML decisions and associated metrics for complementarity. Our proposed framework unifies several previously proposed approaches to combining human-ML decisions. Critically, an analysis of our framework suggests that the optimal mechanism by which human and ML-based judgments should be combined depends upon the specific relative strengths each exhibits in the decision-making application domain at hand. Our optimization setup can be used to generate hypotheses about optimal ways of combining human and ML-based judgments in particular settings, as demonstrated by the simulations in Section~\ref{sec:simulations}. For this, one may use historical decision-making data or models of decision-making for the human and the machine agent. These simulations also help researchers and practitioners understand the trade-offs involved in implementing human-ML collaboration in a decision-making setting by comparing the potential gains in accuracy 
 against the cost of implementation. It is worth noting here that while the joint decision-maker is a theoretical idealized version, in reality the accuracy of the joint decision-maker may be lower due to inefficiencies of real-world decision-making by a human. Thus, it would be useful to quantify the potential benefits of joint decision-making before implementation.  Further, empirically testing the hypotheses and trade-offs presented by our simulations is of great theoretical and practical interest.

Finally, we invite extensions and modifications to our taxonomy, and hope that it serves as a stepping stone toward a theoretical understanding of the broader conditions under which we can and cannot expect human-ML complementarity. 
For example, we invite future research to explore extensions of our proposed optimization problem setup to contexts where predictions do not straightforwardly translate to decisions~\cite{kleinberg2018human}, as well as to settings where the optimal combination of human and ML-based judgment cannot be captured through a convex aggregation function.

\section*{Acknowledgments}
We thank the members of the FEAT ML reading group at Carnegie Mellon University and our anonymous reviewers for their insightful feedback that helped improve this work. During the course of this research, we were supported in part by the UL Research Institutes through the Center for Advancing Safety of Machine Intelligence (CASMI) at Northwestern University. CR was supported partly by the J.P. Morgan AI Research Fellowship, the IBM PhD Fellowship and NSF grant 1763734. LL was supported by the Open Philanthropy AI Fellowship. HH acknowledges support from NSF (IIS2040929 and IIS2229881) and PwC (through the Digital Transformation and Innovation Center at CMU). Any opinions, findings, conclusions, or recommendations expressed in this material are those of the authors and do not reflect the views of NSF or other funding agencies.

\bibliographystyle{apalike}

\bibliography{bibtex}

\end{document}